\documentclass[conference, final]{IEEEtran}

\usepackage{amsmath,amssymb,amsfonts}
\usepackage{algorithmic}
\usepackage{graphicx}
\usepackage{textcomp}
\usepackage{xcolor}
\usepackage{physics}
\usepackage{booktabs}
\usepackage{braket}
\usepackage{microtype}
\usepackage{todonotes}
\usepackage{tikz}
\usetikzlibrary{arrows.meta, positioning}

\usepackage[backend=biber, style=ieee, maxnames=3, minnames=1, url=false, isbn=false, doi=false]{biblatex}
\usepackage{hyperref}
\usepackage{tikz}

\setuptodonotes{inline}

\begin{document}

\title{Entangled Threats: A Unified Kill Chain Model for Quantum Machine Learning Security}

\author{
    \IEEEauthorblockN{
        Pascal Debus\IEEEauthorrefmark{1}, Maximilian Wendlinger\IEEEauthorrefmark{1}, Kilian Tscharke\IEEEauthorrefmark{1}, Daniel Herr\IEEEauthorrefmark{2}, \\
        Cedric Brügmann\IEEEauthorrefmark{2}, Daniel Ohl de Mello\IEEEauthorrefmark{2}, Juris Ulmanis\IEEEauthorrefmark{3}, Alexander Erhard\IEEEauthorrefmark{3}, Arthur Schmidt\IEEEauthorrefmark{4}, Fabian Petsch\IEEEauthorrefmark{4}}
    \IEEEauthorblockA{
    \IEEEauthorrefmark{1}\textit{Fraunhofer Institute for Applied and Integrated Security (AISEC)}, Garching near Munich, Germany\\
    \IEEEauthorrefmark{2}\textit{d-fine GmbH}, Frankfurt, Germany\\
    \IEEEauthorrefmark{3}\textit{Alpine Quantum Technologies (AQT) GmbH}, Innsbruck, Austria\\
    \IEEEauthorrefmark{4}\textit{Federal Office for Information Security (BSI)}, Bonn, Germany
    }
}

\maketitle

\begin{abstract}
Quantum Machine Learning (QML) systems inherit vulnerabilities from classical machine learning while introducing new attack surfaces rooted in the physical and algorithmic layers of quantum computing. Despite a growing body of research on individual attack vectors - ranging from adversarial poisoning and evasion to circuit-level backdoors, side-channel leakage, and model extraction - these threats are often analyzed in isolation, with unrealistic assumptions about attacker capabilities and system environments. This fragmentation hampers the development of effective, holistic defense strategies.

In this work, we argue that QML security requires more structured modeling of the attack surface, capturing not only individual techniques but also their relationships, prerequisites, and potential impact across the QML pipeline. We propose adapting kill chain models, widely used in classical IT and cybersecurity, to the quantum machine learning context. Such models allow for structured reasoning about attacker objectives, capabilities, and possible multi-stage attack paths - spanning reconnaissance, initial access, manipulation, persistence, and exfiltration.

Based on extensive literature analysis, we present a detailed taxonomy of QML attack vectors mapped to corresponding stages in a quantum-aware kill chain framework that is inspired by the MITRE ATLAS for classical machine learning. We highlight interdependencies between physical-level threats (like side-channel leakage and crosstalk faults), data and algorithm manipulation (such as poisoning or circuit backdoors), and privacy attacks (including model extraction and training data inference). This work provides a foundation for more realistic threat modeling and proactive security-in-depth design in the emerging field of quantum machine learning.
\end{abstract}


\section{Introduction}
\label{sec:intro}
As quantum computers become increasingly powerful, there is great interest in finding applications that offer some sort of advantage over classical computers. One promising field is quantum machine learning (QML), which promises many advantages over conventional methods due to the exponentially larger feature space, parameter efficiency, sample complexity, or even robustness. However, new technologies can also pose risks. It is therefore important to examine the security aspects from the outset and to evaluate how QML methods can be used safely. Questions as to how classical development principles (e.g., security by design) can be transferred to quantum computers are therefore of particular interest.

In IT security, it is an established practice to first gain an overview of the available attack surface based on defined protection goals, for example concerning the classical CIA-triad (confidentiality, integrity, availability), and assets worth protecting (data, models, reputation), and to evaluate possible associated attacks to be able to develop structured defense mechanisms. The transition to quantum machine learning results in largely the same protection goals and assets, but the attack surface is significantly larger.

One aspect that is often neglected in current literature on adversarial QML, but is an established process in classic IT security, is threat modeling. A threat model that explains typical threat scenarios and assesses the potential impact according to the attacker's capabilities and resources should be part of any analysis of security aspects of QML. This is the very basis for assessing whether a system is vulnerable and enables better selection and prioritization of defensive measures.

We make the following key contributions:
\begin{itemize}
    \item We present the first kill chain model tailored to QML, adapting established IT security methodologies to the unique properties of QML pipelines.
    \item We provide a process model, inspired by classical killchain models for IT Security and ML, that maps known QML attacks to tactics, techniques, impacted components, and required attacker capabilities.
    \item With this work, we also publish an interactive web application that can be used to explore published adversarial techniques and defenses more conveniently.
    \item Our model highlights the complex interdependencies between physical quantum hardware vulnerabilities, algorithmic manipulation, and privacy attacks.
    \item We propose initial mitigation strategies and identify open research challenges for securing QML systems against multi-stage, realistic threat scenarios.
\end{itemize}

The remainder of this paper is organized as follows:
Section~\ref{sec:related_work} summarizes related work on systemization of knowledge of QML security, while Section~\ref{sec:background} provides some background on typical QML workflows and classical IT security frameworks. Section~\ref{sec:attack_surface} discusses the challenges involved with modeling the QML attack surface, different modeling choices, and their advantages and disadvantages.
The core of the paper consists of Section~\ref{sec:qml_kill_chain_model}, which describes our model of QML-adapted kill chain, and Section~\ref{sec:qml_sec_survey_matrix}, which aligns it with the literature survey we performed. Section~\ref{sec:attack_matrix} introduces our final ATLAS-inspired matrix while Section~\ref{sec:discussion} discusses some findings and discovered gaps in QML research based on our model. Finally, Section~\ref{sec:conclusion} concludes the paper.

\section{Related Work}
\label{sec:related_work}
Security research in quantum machine learning (QML) has progressed alongside the broader field of quantum computing. Most work in this field predominantly analyses very specific vulnerabilities such as crosstalk, side-channel leakage, adversarial evasion, or data poisoning. Investigated attack vectors and defenses range from merely transferring known classical attacks to quantum computers to entirely new quantum-intrinsic attacks, such as noise injection. This variety of attack vectors has already brought about a few \cite{saki_survey_2021,saki_impact_2021,franco_predominant_2024,franco_security_2025}systematization-of-knowledge (SoK) studies and high-level surveys that attempt to catalog known attacks and mitigation strategies.

Most systematization-of-knowledge efforts in QML security revolve around enumerating attacks at the hardware, software, and algorithmic layers. For example, certain works consolidate known side-channel threats like power and timing analysis while separately cataloging algorithmic vulnerabilities such as adversarial examples and data poisoning. Although these SoK studies successfully highlight the breadth of QML security challenges, they often merely present a loosely aligned collection of topics. 

The scope of the different surveys varies greatly. \cite{saki_survey_2021} focuses on security aspects of quantum computing in general, while \cite{saki_impact_2021} discusses noise-related security issues in particular. \cite{franco_predominant_2024} has a clear focus on QML and also performs a prioritization of security aspects. However, vulnerabilities presented in those studies are mostly presented in rather loosely connected categories, with no particular focus on their interconnections. While this categorization helps readers navigate a fast-growing technical field, there is often a lack of threat modeling, attacker taxonomy, and capabilities in the surveys as well as in the literature being surveyed. 
Notably, \cite{franco_security_2025} is not only one of the most extensive surveys but is also the first to suggest a semantic model of the QML attack surface where attack vectors are ordered according to the technology stack of quantum computing and assigned to particular components they are targeting. However, what is also not explicitly discussed in the former study is how multiple vulnerabilities or attack vectors can interact and be utilized by an attacker in a multi-stage attack setting. 
In response to these limitations, our work adopts a kill chain perspective - drawing on established methodologies from classical IT security - to depict how QML attacks can emerge, evolve, and interconnect, which, to the best of our knowledge, has not been done by any other survey or SoK paper yet.

\section{Background}
In this section, we briefly define the scope within the QML field we consider and also provide an overview of the most important components of QML algorithms. We also provide a short introduction to important concepts in classical IT Security and an introduction to so-called kill chain models.

\label{sec:background}
\subsection{The Quantum Machine Learning Pipeline}
Despite a wide variety of proposed QML algorithms, \emph{variational (hybrid) approaches} currently dominate practical implementations and, to some degree, also theory. Early purely quantum algorithms, such as the Harrow–Hassidim–Lloyd (HHL) algorithm for linear systems, remain important theoretical milestones but typically exceed the capabilities of near-term hardware without error correction. In contrast, variational quantum circuits are compatible with so-called Noisy Intermediate-Scale Quantum (NISQ) devices: a comparably short parameterized quantum circuit performs state transformations, and a classical optimizer iteratively refines those parameters based on measurement outcomes. This setup accommodates today’s hardware constraints by limiting circuit depth and qubit requirements.

QML workflows typically begin with \emph{data encoding}, wherein classical inputs (e.g., image pixels, time series values) are represented as quantum states using, for example, angle encoding or amplitude encoding. Next, a \emph{parametric quantum circuit} (often called a variational circuit) processes these states, applying a sequence of gates whose rotation angles serve as trainable parameters. Upon measurement, the circuit’s outputs feed back into a classical optimizer, such as gradient descent, to refine those same parameters over multiple iterations. This cycle constitutes the core workflow for \emph{supervised}, \emph{unsupervised}, and even some \emph{reinforcement learning} QML paradigms. 
In practice, many QML deployments rely on \emph{cloud-based quantum services} that provide the runtime environment, job queue, and error mitigation tools. Jobs submitted through an online interface are \emph{transpiled} to hardware-specific gate sets, then queued and executed on shared quantum processors.

\subsection{Security Concepts and Adversaries}
In classical IT security, defending systems often amounts to preserving the three principal goals \emph{Confidentiality, Integrity, and Availability} (CIA). Confidentiality ensures data, such as proprietary software or sensitive user information, remains accessible only to authorized parties. Integrity requires that data and processes are not tampered with, while Availability demands that legitimate users can reliably access resources. Attackers range from unsophisticated external adversaries testing known exploits to well-funded nation-state actors seeking intellectual property or sabotage. Insiders - such as disgruntled employees or compromised contractors - pose another serious threat due to elevated privileges. Supply chain risks are also an issue when third parties can introduce malicious components into the tech stack.

Although QML deployments add a new type of computing hardware to this picture, the same CIA principles apply. Confidentiality concerns might involve stealing an organization’s proprietary QML model parameters or quantum measurement data, while a successful integrity attack might introduce trojan gates into a victim’s transpiled circuit. Availability threats can manifest in multi-tenant quantum clouds, where a co-tenant floods the hardware with noise-inducing jobs that degrade overall performance.

Therefore, attacker \emph{roles} in a QML context could be further specialized to the following:
\begin{itemize}
    \item \emph{Co-tenant} on a shared quantum machine can exploit crosstalk or observe partial side-channel data.
    \item \emph{Cloud provider insider} may alter transpilation software or calibration routines with malicious intent. 
    \item \emph{Honest-but-curious} third party seeking sensitive information.
\end{itemize}
    
\subsection{Classical Kill Chain Models and Principles}
Beyond the CIA triad, a fundamental cornerstone of modern security strategy is recognizing that threats often unfold in stages. The Lockheed Martin Cyber Kill Chain\footnote{\url{https://www.lockheedmartin.com/en-us/capabilities/cyber/cyber-kill-chain.html}} outlines \emph{reconnaissance, weaponization, delivery, exploitation, installation, command \& control, and actions on objectives}, encouraging defenders to disrupt or detect malicious activity early. 

More recent frameworks like \emph{MITRE ATT\&CK}\footnote{\url{https://attack.mitre.org/}} expand this by categorizing tactics (high-level adversarial objectives) and techniques (specific methods) in a matrix form. This detailed classification has proven especially useful against advanced persistent threats (APTs), providing common terminology and reference scenarios.

Efforts to adapt kill chain modeling to machine learning consider the unique nature of training pipelines, data ingestion, model deployment, and inference APIs. MITRE’s \emph{Adversarial Threat Landscape for Artificial-Intelligence Systems} (ATLAS\footnote{\url{https://atlas.mitre.org/}}) addresses classical ML threats - data poisoning, model evasion, model extraction, etc. - through a stage-based lens. The multi-stage perspective helps defenders see how poisoning might enable stealthy backdoors triggered later, or how reconnaissance queries can precede large-scale model theft. 

As we argue in the next sections, these frameworks can similarly illuminate QML attacks, which combine classical vulnerabilities (e.g., data manipulation) with new quantum attack vectors (e.g., crosstalk, trojans in transpiled circuits).

\section{Decomposition of the QML Attack Surface}
The attack surface of a cyber-physical system, such as a quantum computer made accessible via cloud, is the set of all possible attack vectors that an attacker can use to compromise any of the principal CIA goals. For quantum computing in general and QML in particular, this attack surface is large, complex, and very heterogeneous. In this section, we first illustrate the challenges in modeling this surface and defining its boundaries. We then discuss a variety of possible approaches and conclude by motivating our kill chain approach.

\label{sec:attack_surface}
\subsection{Challenges in Modeling}
Quantum machine learning systems integrate both classical and quantum components, making it difficult to pin down clear security perimeters or even more so to clearly distinguish between classical and quantum parts. In the NISQ regime, data typically flow from a classical pre-processing or feature-engineering pipeline into quantum circuits for partial computation or training updates. Multi-tenant quantum devices further complicate these boundaries, since hardware-level issues like crosstalk and qubit calibration can affect multiple users’ computations in unforeseen ways. Consequently, an attack might originate on the purely classical side, such as feeding poisoned data into the training set, yet ultimately sabotage the quantum model’s reliability. 

\subsubsection{System Boundaries and Hybrid Vectors}
One simple approach to bring order to the variety of attack vectors is to consider from which sources they arise. Firstly, the cloud-based QC-as-a-Service (QaaS) access model leads to a loss of direct control over hardware and server-side software and therefore to supply chain risks and risks relating to third parties. This also implies that QML inherits more or less the full classical attack surface of a typical cloud-deployed (ML-)service or API. Additionally, due to the structure of the QC technology stack (e.g, transpiler, control electronics), which differs from the typical Von Neumann architecture and has not yet been as thoroughly investigated, a new attack surface area emerges from these new components and their interfaces. 

Secondly, focusing more on the software side, it is the objective of many research papers to investigate the transferability of classic ML attack vectors and defenses concerning the special properties of QML. This may result in new attack vectors or attacks that "manifest" differently, but also new QC-intrinsic defense mechanisms. For example, attacks based on quantum noise can be investigated - an intrinsic quantum phenomenon and one of the defining criteria of the NISQ era. 

Finally, there are also hybrid attack vectors that combine or orchestrate classic attacks with QC-specific attacks and blur the line of what is a classical and a quantum attack vector. 

\subsubsection{Criteria for Structured Decomposition}
In the previous section, we have seen that organizing the attack surface in a semantically meaningful way is a challenging task. One approach that was alredy mentioned in the related work section is an \textbf{architecture-oriented decomposition}~\cite{franco_security_2025}, which classifies vulnerabilities with respect to the affected component in the quantum computing tech stack. This perspective is attractive because it allows for assigning each discovered threat, such as side-channel leakage or malicious gate injection, to a specific component. This view is rather static since it ignores the fact that QML model deployment is a process with different stages, such as training and inference, where each stage may also have some characteristic vulnerabilities.

\begin{figure*}
    \centering
    \includegraphics[width=0.75\linewidth]{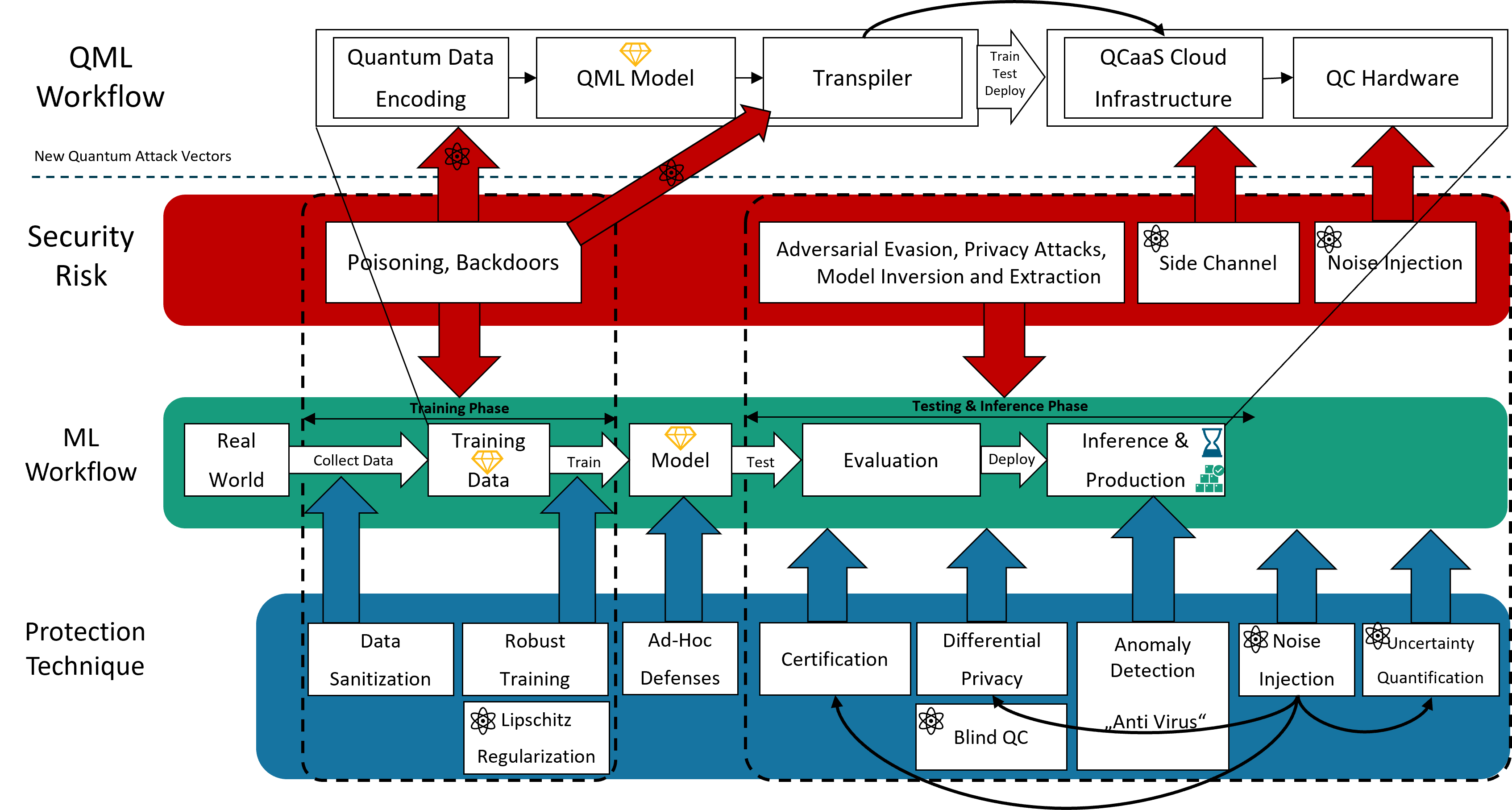}
    \caption{A lifecycle/workflow oriented view on QML}
    \label{fig:qml_workflow}
\end{figure*}

An alternative is consequently a workflow-oriented approach, which considers how QML pipelines unfold from data ingestion and feature engineering, through quantum circuit training and optimization, to final deployment and inference. An illustration of this is shown in Figure~\ref{fig:qml_workflow}. Here, the emphasis is on the practical phases encountered by developers and operators: if an attacker poisons the classical data early in the pipeline, the trained quantum model may learn latent triggers; if the adversary gains access just before deployment, they could embed a backdoor in the transpiled circuit. This mapping can be especially valuable for MLOps teams seeking to pinpoint specific defensive measures, such as data sanitization or circuit verification, at each pipeline stage. Nonetheless, in this view, we still take a developer's perspective, such that defensive measures might not be very well aligned and prioritized according to what an attacker will do.

Finally, a third approach that is frequently employed in classical adversarial modeling is to center on the \textbf{attacker’s perspective} and capabilities. This attacker-centric classification distinguishes, for instance, an external black-box adversary probing an API from a co-tenant who can access shared qubits and glean side-channel leaks, or a malicious cloud provider that wields complete control over the transpiler and scheduling layers. By exploring adversaries’ privileges, motivations, and potential targets, this perspective shows which exploits are plausible or cost-effective. Yet it can fracture the overarching picture into separate attacker profiles without clarifying how a multi-stage campaign might progress - from reconnaissance, to initial foothold, to circuit-level exploitation.

\subsection{Evaluating Decomposition Strategies}
These three presented viewpoints each illuminate certain aspects of the QML attack surface while obscuring or oversimplifying others. Architecture-based methods excel at showing where a threat physically resides, an advantage when diagnosing software or hardware vulnerabilities that could otherwise be overlooked. Workflow-based taxonomies are more natural for developers and operators, who tend to see QML systems as sequences of discrete stages from data intake to circuit execution. Meanwhile, focusing on adversary roles underscores realistic threat scenarios tied to user privileges, especially in multi-tenant or cloud-based quantum computing contexts. Yet, if the goal is to understand how multiple vulnerabilities can be chained together - and to determine at which point a defender can most effectively intervene - none of these decompositions alone suffices.

Workflow-based defenses might protect only discrete steps, missing the fact that a malicious co-tenant with scheduling influence can reinsert noise-based attacks throughout the model’s lifecycle. Similarly, an attacker-centric approach, while clarifying who might attack and why, will not inherently show \emph{when} an attacker’s techniques are best deployed. In short, each method has strengths for enumerating or classifying threats, yet they do not, in isolation, capture the evolving, multi-stage nature of modern adversarial operations.

The complexity and fluidity of QML threats argue strongly for a kill chain model, which maps adversarial actions onto a structured, chronological progression from reconnaissance to exploitation and beyond. By incorporating aspects of architecture-based, pipeline-based, and attacker-centric thinking, a kill chain broadens the traditional concept of “attack surface” into an “attack lifecycle.” 

\subsection{Motivation for a Kill Chain Perspective}
First, a kill chain model naturally supports defense in depth, since it shows how an attacker’s early steps, such as side-channel reconnaissance or partial data poisoning, can be disrupted before they progress to more damaging stages like model theft or circuit backdoors. Defenders can tailor mitigations to each phase of the lifecycle, enabling them to stop attacks earlier and reduce the chances of full system compromise. Second, the kill chain perspective clarifies that certain attack techniques often appear at multiple stages, or become intertwined in ways that neither an architecture-based nor pipeline-based view alone might reveal. For instance, noise-injection attacks apply both to training (to degrade model accuracy) and inference (to force targeted misclassifications).

Additionally, a lifecycle approach highlights how some attacks, particularly hybrid vectors, depend on knowledge or access obtained in a prior phase. An attacker who uses side channel analysis to map circuit structure can later mount a more tailored Trojan injection via the transpiler. Recognizing this sequential chain of dependencies ensures that defenders understand the bigger picture: a single infiltration step can fuel a cascade of interlinked exploits. Equally important, kill chains have a proven track record in classical cybersecurity, helping organizations see the progression of adversarial campaigns and prioritize resources to disrupt attackers early in the chain. In QML contexts, where the interplay of quantum hardware, specialized compilers, and multi-tenant cloud infrastructures presents novel opportunities for stealth and escalation, the kill chain paradigm offers a unifying framework that is far more robust than enumerating isolated threats.

In the subsequent section, we build upon these observations to propose a QML-specific kill chain, aligning hardware, pipeline, and attacker-centric insights into a stage-based model. By adapting familiar classical kill chain principles to quantum, we aim to provide a clearer roadmap of how adversaries can combine multiple exploits and where defenders can most effectively intervene to safeguard confidentiality, integrity, and availability in QML systems.

\section{A Kill Chain Model for QML}
\label{sec:qml_kill_chain_model}
In this section, we introduce our kill chain model by first defining a subset of relevant stages and subsequently defining what set of key fields or attributes (attacker role, Prerequisites, etc.) we add to the description of individual techniques or subtechniques.

\subsection{A Simplified Stages Model}
\begin{figure*}[h]
\centering
\begin{tikzpicture}[
  node distance=0.4cm,
  every node/.style={draw, rectangle, rounded corners, minimum height=1.2cm, minimum width=2.8cm, align=center},
  arrow/.style={thick,->,>=Stealth}
]
\node (Recon) {Reconnaissance};
\node (Access) [right=of Recon] {Initial Access};
\node (Manipulate) [right=of Access] {Model Manipulation};
\node (Persist) [right=of Manipulate] {Persistence};
\node (Exfil) [right=of Persist] {Exfiltration / Impact};

\draw[arrow] (Recon) -- (Access);
\draw[arrow] (Access) -- (Manipulate);
\draw[arrow] (Manipulate) -- (Persist);
\draw[arrow] (Persist) -- (Exfil);
\end{tikzpicture}
\caption{Quantum Machine Learning Kill Chain Stages}
\label{fig:killchain}
\end{figure*}
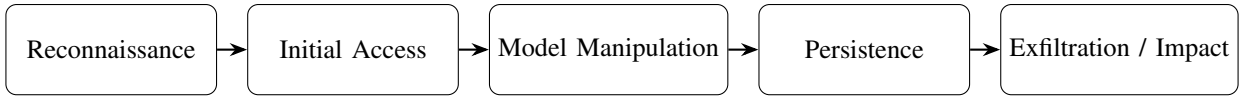

Drawing on classical kill chain frameworks like Lockheed Martin’s Cyber Kill Chain and MITRE ATT\&CK, we adapt a five-stage model to accommodate quantum machine learning’s unique architecture and lifecycle. Our stages - \emph{Reconnaissance}, \emph{Initial Access}, \emph{Model Access / Manipulation}, \emph{Persistence}, and \emph{Exfiltration / Impact} - cover the broad sweep from an adversary’s early information gathering through final destructive or exfiltrative actions. As shown in Figure~\ref{fig:killchain}, each stage builds on potential successes in earlier phases.

\subsubsection{Reconnaissance Stage}
The reconnaissance stage encompasses both general discovery techniques and quantum-specific methods for gleaning circuit structure or resource usage. In classical IT systems, reconnaissance typically involves scanning for open ports or unpatched services, whereas QML contexts introduce subtler vectors like power or timing side-channels that may reveal information about circuit structure and even encoded data.

\subsubsection{Initial Access Stage}
Once the attacker identifies a workable entry point, they move to the initial access stage, wherein they gain partial but meaningful footholds. This access can manifest as manipulating scheduling algorithms that allocate jobs for securing concurrent execution with victim circuits on a cloud quantum device, or injecting malicious data into a shared dataset.

\subsubsection{Model Access / Manipulation Stage}
With a foothold established, this stage marks the pivot to actively interacting with the victim’s QML model or associated circuits. An adversary might tamper with parameterized gates to degrade accuracy, embed hidden trojans in the transpiled circuit, or orchestrate additional noise injections during training. This manipulation can occur throughout the iterative training loop, particularly in variational circuits, and may remain stealthy if the attacker selectively triggers anomalies.

\subsubsection{Persistence Stage}
Depending on the overall goal, an attacker might also try to implement stealthy long-term control mechanisms in the system or the victim model. The most direct example of this is Trojan attacks via data poisoning, but also other possibilities, like deliberate miscalibration for gate injection, might be possible here.

\subsubsection{Exfiltration / Impact}
Finally, the last stages capture the culmination of adversarial goals. In some cases, exfiltration involves stealing quantum model parameters or measurement data, effectively replicating proprietary QML logic without paying for the model’s development. Other times, the primary objective is sabotage, such as forcing misclassifications, causing DoS conditions via amplified noise, or revealing private data used in training. This concluding stage underscores that, while multi-stage attacks can manifest at any point, the attacker’s ultimate aim is typically realized when they either exfiltrate sensitive assets (e.g., trained QML models, sensitive data) or produce disruptive impacts (e.g., inference failures).

Compared to a full MITRE ATT\&CK or ATLAS matrix, we omit phases like “weaponization” or “command-and-control,” as quantum jobs and transpiler modifications generally skip the classical concept of delivering malicious binaries or creating remote access trojans. Nonetheless, the spirit of a kill chain remains: each stage is an opportunity for defenders to detect and thwart the adversary, enabling “defense in depth” by layering protective measures at multiple points.

\subsection{Definition of Techniques}
Within each kill chain stage, adversaries can employ \emph{techniques} that often map to concrete attacks published in QML security research. For example, the “crosstalk-based denial-of-service” technique described in the literature fits very well into the Impact stage, once the attacker has gained initial access to a multi-tenant quantum device. Conversely, “power side-channel leakage” is a typical reconnaissance technique that reveals gate sequences or parameterization details before the adversary tries to manipulate the circuit execution in subsequent stages.

We formalize each technique by specifying crucial attributes essential to capturing a multi-stage adversarial campaign. These include the \emph{attacker role} (e.g., co-tenant, malicious provider, or external black-box user) and the \emph{capabilities} required to execute the technique, such as direct control over the transpiler or ability to run concurrent jobs on adjacent qubits. We then note the \emph{Prerequisites}, for instance, needing partial knowledge of the victim’s circuit layout or scheduling times, to situate the technique realistically within the kill chain. Next, we list \emph{impacted components}, clarifying whether the exploit compromises circuit parameters, data encoding, measurement readout, or the classical code that optimizes the circuit’s parameters. Finally, we identify \emph{possible defenses} and mitigation approaches - for instance, verified compilation can detect malicious gates, differential privacy can mitigate data poisoning, and hardware isolation can prevent crosstalk interference.

This technical detail is critical because it closes the loop from abstract stage definitions to tangible, system-level interventions. By bringing in role-specific capabilities, we show that co-tenants with access to an apparently benign calibration job can actually break a victim’s model, whereas a strictly external adversary might lack that path. Just as importantly, clarifying prerequisites reveals how certain techniques interlock; a side-channel technique in reconnaissance may feed precisely the knowledge an attacker needs for a high-impact manipulation in the exploitation phase. In this way, the kill chain perspective not only enumerates attacks but emphasizes the relationships among them, giving defenders actionable insights into how to detect, disrupt, or mitigate adversarial campaigns before they escalate into full compromise.

\section{A Kill Chain-oriented Literature Review}
\label{sec:qml_sec_survey_matrix}
In the following section, we provide a kill chain-oriented literature review. Since some of the techniques or their respective subtechniques or variations presented in the literature apply to multiple stages, we present our literature review grouped by techniques. This also better aligns with the perspective that most readers are more familiar with. For each technique, we provide a short description, summarize QML-related publications with references, motivate a stage assignment with our kill chain model, and provide key fields with threat-related meta-information.

\subsection{Side-channel attacks (SCA)} 
One of the main targets of SCA is reverse engineering (RE) of the quantum circuit. Since data is also encoded as a special embedding circuit, an attacker can steal both the QML model and the training or inference data. SCA based on powertraces, which can be calculated directly from the control electronics or the pulse program, have already been described in the literature \cite{erata_quantum_2024,xu_exploration_2023} or based on timing \cite{lu_quantum_2024} and successfully tested on IBM QC hardware. Most recent publications also use SWAP gates\cite{lee_swap_2025} or crosstalk \cite{choudhury_crosstalk-induced_2024} for side-channel attacks. Depending on the attacker's access (varying from concurrent user to hardware access), different information about the circuit can be extracted up to full RE. 

\begin{itemize}
    \item \textbf{Stage Assignment}: Side-channel techniques are typically associated with Stage 1 (Reconnaissance), where an attacker gathers critical information from power traces, timing, or other physical signals. Ultimately, SCA can feed into Stage 5 (Impact) if the attacker aims to exfiltrate circuit details or data.
    \item \textbf{Attacker Role}: Co-tenant user with partial hardware access, or malicious insider with physical or firmware-level observation.
    \item \textbf{Attacker Capabilities}: Measuring power usage, gate timing, or residual states in multi-tenant hardware. Possibly requires specialized instrumentation or job co-location.
    \item \textbf{Prerequisites}: Enough overlap or scheduling concurrency to eavesdrop on the victim’s run. In some cases, partial knowledge of the circuit gating patterns is required.
    \item \textbf{Impacted Components}: model, data
    \item \textbf{Possible Defenses}: Noise-shielding or random delays in scheduling, hardware obfuscation (randomization of control pulses), isolation of qubit groups across tenants, verified run logs.
\end{itemize}

\subsection{Poisoning} 
Poisoning attacks are about manipulating the training data of a (Q)ML model with the aim of either degrading the performance of the model as a whole, possibly to the point of uselessness (attack on availability), or specifically, i.e., only for different pairs of classes (attack on integrity). Trojan attacks are a special case \cite{huang_survey_2020,liu_trojaning_nodate} in which misclassification is only triggered in the presence of an (invisible) watermark introduced by the attacker. Data can be manipulated using a spectrum of methods, starting with simple Gaussian noise on numerical features \cite{xiao_is_2015}, randomized or targeted permutation and flips of categorical features or labels \cite{xiao_adversarial_2012} up to adversarial noise as a special form of the aforementioned watermarks.  
\begin{itemize}
    \item \textbf{Stage Assignment}: Poisoning usually arises in Stage 2 (Initial Access) when the attacker manipulates training data before or during model construction
    \item \textbf{Attacker Role}: Often a data supplier or user who can contribute or alter training data; in some cases, an untrusted service provider with write access to the data pipeline. 
    \item \textbf{Attacker Capabilities}: Ability to insert or modify training samples, labels, or data-encoding gates. Possibly also knowledge of the model’s structure to determine the best gate insertion points
    \item \textbf{Prerequisites}: Access to or influence over the data ingestion phase, or partial knowledge of how data is encoded into quantum circuits.
    \item \textbf{Impacted Components}: Training dataset, Model availability/integrity.
    \item \textbf{Possible Defenses}: Outlier/consistency checks, differential privacy to limit individual sample influence, robust training strategies (e.g., adversarial training on classical + quantum data), verified circuit compilation
\end{itemize}

\subsection{Evasion} Adversarial evasion attacks are characterized by the fact that they deliberately but minimally perturb the input of a QML model to achieve an incorrect classification by the model. A variety of different attacks have been researched in the classical case, and the vulnerability of QML methods was established (in theory) early on by \cite{liu_vulnerability_2020}.

In empirical studies, vulnerability to standard white box methods (Fast Gradient Sign Method\cite{goodfellow_explaining_2015}, Projected Gradient descent \cite{madry_towards_2019}) in the simulator \cite{lu_quantum_2020} and also on real quantum hardware \cite{ren_experimental_2022} for both classical and quantum data was shown in various works. Simple defenses such as adversarial training have also been evaluated (positively). \cite{west_benchmarking_2023} also investigates transfer attacks as gray-box/black-box methods and suggests a possible quantum advantage in robustness. Adversarial robustness, as well as the influence of data encoding and advanced defense mechanisms such as Lipschitz regularization, was further investigated by \cite{wendlinger_comparative_2024} and led to some limitations of this quantum advantage claim.

Defense techniques, inspired by many classical noise-based defenses, mainly investigate the influence and use of quantum noise to improve robustness \cite{du_quantum_2021,west_drastic_2024}. However, there are also more innovative approaches \cite{gong_enhancing_2024} that use randomized encoding and decoding circuits to create a "barren plateau" for the attacker \cite{mcclean_barren_2018}, which makes the generation of adversarial samples more difficult. Furthermore, at the interface to robustness certification, the influence of Lipschitz regularization was investigated by \cite{berberich_training_2023} and by  \cite{wendlinger_comparative_2024}. 

Finally, various classic techniques for (formal) verification are transferred to QML models. These are based, for example, on the calculation of error bounds using classical optimization methods such as Semidefinite Programming (SDP), as described in \cite{guan_robustness_2021} and/or on randomized smoothing \cite{weber_optimal_2021}.

\begin{itemize}
    \item \textbf{Stage Assignment}: Evasion-based adversarial examples typically occur during Stage 3 (Model Access/Manipulation) if the attacker can probe the model and craft inputs to cause misclassification. Alternatively, the actual exploit - submitting adversarial inputs - can appear at Stage 5 (Exfiltration/Impact) if the attacker’s goal is immediate manipulation, e.g., forcing a classifier to produce incorrect results in a critical application.
    \item \textbf{Attacker Role}: External user with black-box or gray-box query access, or insider with detailed model knowledge (white-box scenario).
    \item \textbf{Attacker Capabilities}: Ability to supply or modify inference inputs. Potentially uses classical methods (FGSM, PGD) adapted for quantum data encoding.
    \item \textbf{Prerequisites}: Sufficient knowledge of how the QML model processes data (e.g., gradient or surrogate model in black-box settings).
    \item \textbf{Impacted Components}: Model’s inference pipeline, user-facing prediction APIs
    \item \textbf{Possible Defenses}: Adversarial training with quantum or classical techniques, noise-based robustness improvements \cite{du_quantum_2021,west_drastic_2024}, randomized encoding/decoding \cite{gong_enhancing_2024}, Lipschitz regularization \cite{berberich_training_2023,wendlinger_comparative_2024}.
\end{itemize}

\subsection{Noise Attacks} 
Current NISQ hardware is in a fragile balance between isolation from disturbing environmental influences and sufficient coupling to the environment to enable effective (optimal) control. Any disturbance of this balance or even sub-optimal control, especially intentional, leads to more or less strong noise in the calculation and thus incorrect results. Direct influence by third parties on calibration or mapping data, as described by \cite{ash-saki_analysis_2020}, must therefore obviously lead to problems. More subtle and realistic are cross-talk induced attacks in co-tenancy mode, either through CNOT drives of neighboring qubits \cite{ash-saki_analysis_2020, harper_crosstalk_2024} or excessive shuffling in ion traps \cite{saki_shuttle-exploiting_2021}. Defenses have also already been proposed in this direction: anti-virus patterns \cite{deshpande_towards_2022}, matching for the detection of interference circuits and buffer qubits \cite{sadi_special_2022}. In contrast to glitching/fault injection in classic hardware security, all attacks proposed in this direction to date have been aimed more at a deterioration in performance (i.e. an attack on availability) rather than at targeted manipulation of the results (e.g. changing a classification result) in the sense of an attack on integrity.

\begin{itemize} 
    \item \textbf{Stage Assignment}: Intentional noise injection is typically a form of Stage 3 (Model Manipulation). It might also manifest in Stage 5 (Impact), particularly if the attacker’s explicit goal is denial-of-service or forced misclassification.
    \item \textbf{Attacker Role}: Co-tenant who can induce crosstalk, malicious cloud operator adjusting calibration or scheduling.
    \item \textbf{Attacker Capabilities}: Physical or software-level adjustments to qubit operations, pulse sequences, or concurrent job scheduling.
    \item \textbf{Prerequisites}: Access to hardware-level parameters or co-located qubits for crosstalk-based sabotage.
    \item \textbf{Impacted Components}: QML training (degraded convergence), performance (higher misclassification rates)
    \item \textbf{Possible Defenses}: Run-time circuit verification, job isolation, “antivirus” patterns that detect or neutralize abnormal gates \cite{deshpande_towards_2022}, matching/buffer qubits \cite{sadi_special_2022}.
\end{itemize}

\subsection{Measurement} 
The situation is very similar concerning availability attacks on the measurement process. Shot noise is already a source of error without deliberate manipulation. Any manipulation at the level of the circuits, the pulse sequence, or the measurement parameters will amplify this effect. Again, there are no targeted attacks on availability yet, but the attack by \cite{saki2021qubitsensingnewattack} is worth mentioning, in which potentially sensitive output information can be extracted from the measurement process.

\begin{itemize} 
    \item \textbf{Stage Assignment}: Attacks on the measurement process typically appear in Stage 3 (Model Manipulation) if the adversary modifies shot-taking parameters or readout pulses to skew results, or in Stage 5 (Impact) if the ultimate effect is either data theft (reading sensitive states) or denial-of-service (making measurement unusable). 
    \item \textbf{Attacker Role}: Cloud provider insider adjusting calibration, or co-tenant interfering with the circuit before measurement.
    \item \textbf{Attacker Capabilities}: Access to measurement parameters or the ability to run concurrent noisy measurement routines.
    \item \textbf{Prerequisites}: Overlapping or high-privilege control of qubit readouts, scheduling, or calibration data.
    \item \textbf{Impacted Components}:  model integrity, model ouputs
    \item \textbf{Possible Defenses}: Verified measurement processes, improved shot noise reduction, circuit-level verification, 
\end{itemize}

\subsection{Backdoors} The goals of backdoors are either degrading the performance (attack on availability) or targeted manipulation (attack on integrity) of the model based hidden triggers in the inputs. Since QML data in particular also corresponds directly to parts of the circuit, gate injection is the obvious attack vector for a backdoor \cite{chu_qtrojan_2023}. Depending on the attacker's access capabilities, this can either be done directly or more covertly within the toolchain, such as with approximate transpilation/synthesis \cite{chu_qdoor_2023} of the final circuit. 
There are tools for circuit equivalence checking \cite{burgholzer_advanced_2021,burgholzer_verifying_2020,peham_equivalence_2022,sander_equivalence_2024} that can be used for defense, but these can only be used if the transpilation can be carried out locally under the control of the user.

\begin{itemize} 
    \item \textbf{Stage Assignment}: Backdoors fit best in Stage 4 (Persistence) when embedded into training data and injected either directly or via the transpiler.
    \item \textbf{Attacker Role}: Malicious provider controlling transpiler synthesis \cite{chu_qdoor_2023}, or external adversary who hijacks training data.
    \item \textbf{Attacker Capabilities}: Ability to insert or replace gates, approximate circuit synthesis, or embed triggers in data.
    \item \textbf{Prerequisites}: Knowledge of how the user trains or deploys the QML model, or privileged compiler-level access.
    \item \textbf{Impacted Components}: Model integrity, final classification results
    \item \textbf{Possible Defenses}: Circuit equivalence checking \cite{burgholzer_advanced_2021,peham_equivalence_2022}, transpiler audits, restricting approximate circuit synthesis to known libraries, and local verification of final circuits.
\end{itemize}

\subsection{Model stealing} The goal of model stealing is basically reverse engineering (RE) of a victim model based on the access and information available to an attacker. This implies that sometimes only rough architecture parameters (number/type of layers) can be stolen, while sometimes it is possible to exactly replicate the victim model. Often, it is not necessary to copy the model exactly, but only to approximate its behavior well enough. In this case, an attacker would first try to extract the architecture and then train a similar model, possibly using the original model as a "teacher" via query access. Furthermore, a distinction can be made between classic attacks/defenses and QML/QC-specific ones. 

An obvious (classical) attack vector is direct copying of the model by third parties (hardware providers, QC software manufacturers): In the quantum as a service model (QaaS) via cloud and the associated supply chain of third-party components (transpilers, compilers, control electronics, etc.),  there are a large number of possibilities for this, as  \cite{ash-saki_analysis_2020} compactly summarized. Much of the research on QML model stealing is therefore focused on various more or less QML-specific defense strategies, such as reinforcement learning-based distributed model architectures \cite{wang_qumos_2023},  insertion of dummy swap gates \cite{suresh_short_2021}. A more "traditional" extraction in the sense of model approximation is discussed in \cite{fu_quantumleak_2024}.

\begin{itemize}
    \item \textbf{Stage Assignment}: Model stealing can unfold across multiple stages but often culminates in Stage 5 (Exfiltration/Impact) if the attacker’s ultimate goal is to replicate the victim’s QML model. The actual extraction may require Stage 3 (Model Access) (e.g., repeated queries)
    \item \textbf{Attacker Role}: Often a malicious service provider (full hardware/software access) or an external adversary with black-box query privileges.
    \item \textbf{Attacker Capabilities}: Potentially extensive, from simply reading the transpiler’s output to controlling the entire quantum cloud environment.
    \item \textbf{Prerequisites}: In black-box scenarios, enough queries to approximate the model. In insider scenarios, direct readout of circuit instructions or measurement results.
    \item \textbf{Impacted Components}: Model IP (architecture, parameters), data
    \item \textbf{Possible Defenses}: Encrypted or blind quantum computation, distributing the QML model to multiple parties so no single provider sees all aspects, insertion of dummy SWAP gates or watermarks \cite{suresh_short_2021}, verified compilation.
\end{itemize}

\subsection{Privacy} As in the case of classical ML, a privacy attack in QML aims to extract information about training data. Depending on the scope of the possible extraction, there is a spectrum between complete reconstruction or inversion attacks \cite{fredrikson_model_2015} as well as so-called membership or property inference \cite{choquette-choo_label-only_2021} in which only individual properties or affiliations to (sub)sets of the data set can be found for individual data points. The standard countermeasure to this is the concept of differential privacy (DP), in which the amount of extractable information can be limited by targeted noise in the data, target function, or outputs.

Most of the research published on QML is theoretical within the framework of the formal $\delta$,$\epsilon$-DP definition and transfers various DP concepts to QC and QML. \cite{yang_improved_2023, zhou_differential_2017} investigates DP with different types of quantum noise, \cite{hirche_quantum_2023} provides upper bounds for privacy parameters using an information-theoretic interpretation. \cite{heredge_prospects_2024} is the only paper that explicitly deals with the conditions for membership inference and inversion attacks (weak and strong privacy breaches) via gradients outside the formal DP definition. On the empirical side, \cite{guan_detecting_2023} specializes in detection algorithms for DP violations, and \cite{watkins_quantum_2023} trains a DP-QML model with VQC architecture that demonstrates comparable or better performance than a classic baseline model with the same privacy budget. In addition to DP, the entire field of confidential or blind QC also opens up as a defence technique, but we will skip this at this point, as it is not QML-specific and is still a long way from technical implementation.

\begin{itemize}
    \item \textbf{Stage Assignment}: Privacy attacks generally surface in Stage 5 (Exfiltration/Impact), once the adversary aims to recover sensitive training data or membership information.
    \item \textbf{Attacker Role}: Black-Box Attacker with query access, Co-tenant user with victim circuits
    \item \textbf{Attacker Capabilities}: Ability to query or inspect model outputs, gradient updates, or raw measurement results.
    \item \textbf{Prerequisites}: Either interactive query capability (like membership inference) or direct access to a snapshot of the QML training process.
    \item \textbf{Impacted Components}: Confidentiality of training data, model outputs, gradient updates.
    \item \textbf{Possible Defenses}: Differential privacy (targeted noise injection), restricting model output granularity, secure enclaves for quantum jobs, blind quantum computing.
\end{itemize}

\section{An (ATLAS-inspired) Attack Matrix}
\label{sec:attack_matrix}
Following the example of the MITRE ATLAS, we present the results of the previous section~\ref{sec:qml_sec_survey_matrix}  more compactly in the form of a matrix. For this work, we provide a high-level overview with all important key fields on attacker capabilities, prerequisites, and affected components in Table~\ref{tab:qc_techniques}. Here, we focus on a lower granularity view of the attack vectors at the level of techniques, as presented in the previous section, and provide the references to publications that discuss either attack or defense \emph{subtechniques} related to a particular technique.

\begin{table*}[htb]
\centering
\caption{Overview of QML Attack Techniques and Stages in the Proposed Kill Chain.}
\label{tab:qc_techniques}
\begin{tabular}{p{1.5cm} p{1.5cm} p{2.3cm} p{2.3cm} p{2.3cm} p{2.3cm} p{1.0cm} p{2.3cm}}
\hline
\textbf{Technique} & \textbf{Stage} & \textbf{Attacker Role} & \textbf{Capabilities} & \textbf{Prerequisites} & \textbf{Component} & \textbf{References} & \textbf{Defenses} \\ 
\hline

\textbf{SCA}
& \textbf{Stage 1 (Reconnaissance)}
& Co-tenant, malicious insider 
& Observe power/timing/pulse usage
& Overlapping job scheduling or firmware-level access
& Circuit structure, Data, Model
& \cite{erata_quantum_2024, xu_exploration_2023, lu_quantum_2024}
& Noise-shielding, random scheduling, isolation \\[6pt]

\textbf{Poisoning} 
& \textbf{Stage 2 (Initial Access)} 
& Data supplier, co-tenant 
& Insert or modify training data, embed trojans 
& Ability to influence data ingestion or quantum encoding 
& Training set, model parameters (availability/integrity) 
& \cite{xiao_is_2015, xiao_adversarial_2012, huang_survey_2020, liu_trojaning_nodate, chu_qtrojan_2023, chu_qdoor_2023}
& Outlier checks, DP, robust/ adversarial training \\[6pt]

\textbf{Evasion}
& \textbf{Stage 3 (Model Access / Manipulation)}
& External/partial insider with query access
& Craft adversarial inputs (e.g., FGSM, PGD) 
& Sufficient model knowledge or iterative queries
& Model inference, quantum data encoding
& \cite{liu_vulnerability_2020, gong_universal_2021, lu_quantum_2020, ren_experimental_2022, west_benchmarking_2023}
& Adversarial training \cite{du_quantum_2021}, noise-based defenses \cite{gong_enhancing_2024}, Lipschitz reg. \cite{berberich_training_2023, wendlinger_comparative_2024} \\[6pt]

\textbf{Noise Attacks}
& \textbf{Stage 3 (Model Access / Manipulation)}
& Co-tenant or cloud operator 
& Inject crosstalk/noise or tamper with calibrations
& Scheduling concurrency or access to calibrations
& Model accuracy, training/inference fidelity
& \cite{ash-saki_analysis_2020, harper_crosstalk_2024, saki_shuttle-exploiting_2021}
& Job isolation, antivirus patterns \cite{deshpande_towards_2022}, buffer qubits \cite{sadi_special_2022} \\[6pt]

\textbf{Backdoors}
& \textbf{Stage 4 (Persistence)}
& Malicious provider, external attacker w. transpiler access
& Insert hidden gates, approximate synthesis trojans
& Compiler-level control or data injection 
& Model integrity, final classification logic
& \cite{chu_qtrojan_2023, chu_qdoor_2023}
& Circuit equivalence checks \cite{burgholzer_advanced_2021, peham_equivalence_2022}, local transpilation \\

\textbf{Privacy}
& \textbf{Stage 5 (Exfiltration / Impact)}
& Cloud provider (honest-but-curious), co-tenant 
& Read model outputs/gradients or measurement logs
& Query interface or partial hardware/log access
& Confidentiality of training data, encoded states
& \cite{fredrikson_model_2015, choquette-choo_label-only_2021, heredge_prospects_2024}
& Differential privacy \cite{yang_improved_2023, zhou_differential_2017, watkins_quantum_2023}, restricted output granularity \\[6pt]

\textbf{Model Stealing}
& \textbf{Stage 5 (Exfiltration / Impact)}
& Malicious provider, external black-box attacker
& Clone or approximate QML model parameters/structure
& Query-based learning or direct circuit readout
& Model IP, quantum circuit layout, training pipeline
& \cite{fu_quantumleak_2024, ash-saki_analysis_2020}
& Blind/confidential QC, circuit watermarking \cite{suresh_short_2021}, distributed QML \cite{wang_qumos_2023} \\[6pt]

\textbf{Measurement Attacks}
& \textbf{Stage 5 (Exfiltration / Impact)}
& Cloud provider insider, co-tenant w. partial control
& Manipulate readout pulses/parameters 
& Physical or job-level influence on measurement process
& Availability of correct results, potential confidentiality leak
& \cite{saki2021qubitsensingnewattack} 
& Verified measurement procedures, shot noise suppression \\[6pt]

\hline
\end{tabular}
\end{table*}

\begin{figure*}
    \centering
    \includegraphics[width=\linewidth]{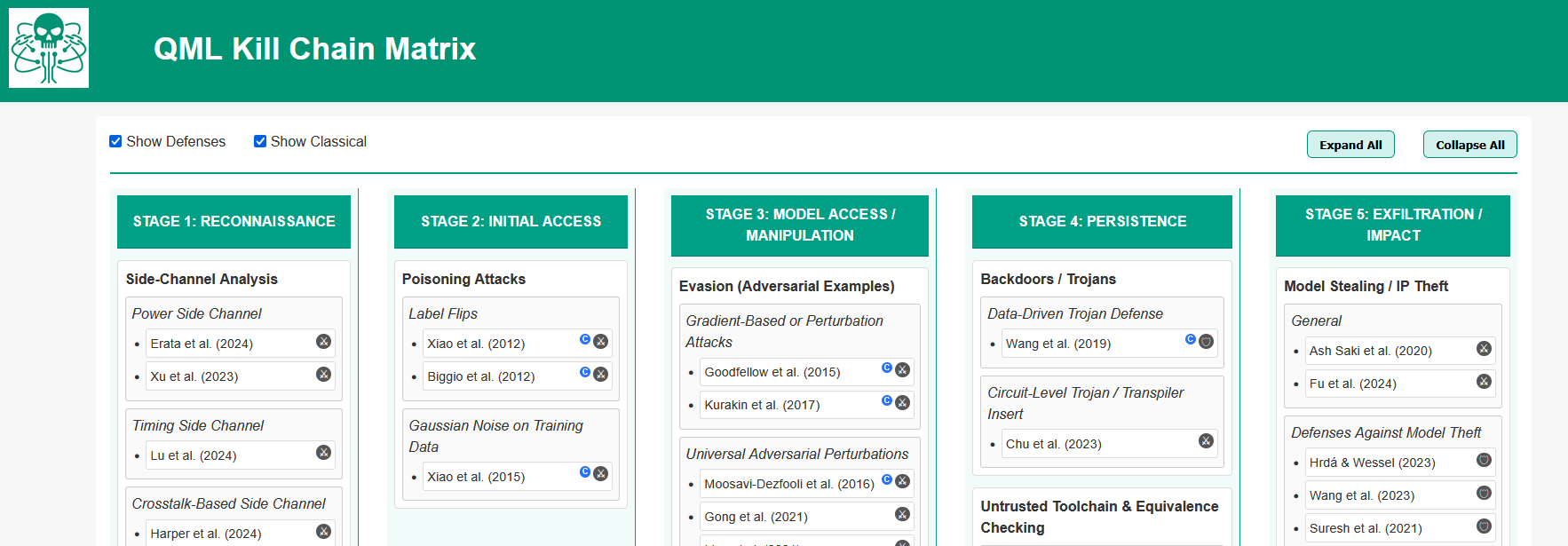}
    \caption{Illustration of the interactive QML Killchain Web Application}
    \label{fig:killchain_app}
\end{figure*}

Since a more detailed, higher granularity view on the techniques is beyond the scope of this work, we also created an interactive web application of our kill chain model, where users can find more detailed threat-related information, also at the subtechnique level, as well as corresponding literature references. A screenshot of the first version of this application\footnote{\url{https://debus.pages.fraunhofer.de/qmlkillchain/}} is shown in Figure~\ref{fig:killchain_app} and will be finalized upon publication.

\section{Discussion and Future Directions}
\label{sec:discussion}

\subsection{Findings from the Kill Chain Modeling}
The QML kill chain model underscores the inherent complexity and multi-layered nature of quantum adversarial threats. Our survey in Section 6 highlights that techniques once confined to classical ML - such as data poisoning or model extraction - take on new dimensions in a QML context, where hardware-level noise and quantum-circuit encodings open further vulnerabilities. For example, side-channel reconnaissance at Stage 1 can yield crucial insights (e.g., qubit connectivity, gate parameters) that enable malicious gate modifications at Stage 3 or injection of a stealthy backdoor at Stage 4. By mapping specific attacks to kill chain stages, we see more clearly that many partial solutions proposed in the literature do not account for these inter-stage synergies.

The kill chain approach thus reinforces the defense in depth principle by pinpointing opportunities to disrupt adversaries early. If crosstalk-based side-channel leaks are minimized at Stage 1 through hardware isolation or random scheduling, subsequent phases - like targeted circuit modification - become significantly harder. Similarly, verifying compiled circuits for hidden trojans before deployment can thwart persistent backdoors that might remain dormant for extended periods.

\subsection{Identified Gaps in Literature and Practice}
When analyzing the techniques published in the literature to assign them to the stages or define the key fields for a technique, it becomes apparent that many papers lack a formal threat model and often a clear presentation and discussion of the underlying assumptions, such as attacker capabilities or required access level.  As a result also most QML security surveys do not comment on the execution complexity of a particular attack vector and treat each threat or defense in isolation.

If information on attacker roles and capabilities were more explicitly discussed in current and future QML security publications, it could be a major contribution also for the prioritization of research on defensive techniques, since it becomes clearer how likely a particular attack is going to occur. An important second building block in this direction, which is also a gap in the current literature, is the exploration of multi-stage attacks and the development of corresponding proof-of-concept realizations of such attacks. An example of this was already mentioned at the beginning of this section.

On the level of individual techniques, adversarial evasion is a very dominant field, nevertheless, it clusters around works focusing \emph{in depth} on simple adversarial sample generation and the effect of noise, either simulated or on real hardware. Being one of the most important fields also in classical adversarial machine learning, we argue that it should also be considered more in breadth. Interesting new research directions, that have only been initially explored~\cite{wendlinger_comparative_2024,west_benchmarking_2023}, are transfer attacks from classical models (and how to avoid them) as well as Lipschitz-constant-based regularization~\cite{berberich_training_2023}, which is computationally not feasible for classical models. Furthermore, many discussed techniques, especially for noise injection, focus on attacking the availability of model by just degrading performance while attacks on model integrity by realizing precicesly targeted perturbations are not often realized.

\subsection{Prospects for Future Research}
QML’s near-term future includes a transition from noisy intermediate-scale devices to (eventually) fault-tolerant quantum computers. It remains unclear which vulnerabilities will persist or morph once powerful error-correction is in place. For instance, crosstalk-based manipulation might become less relevant if error-correcting codes successfully mask hardware-level perturbations. On the other hand, error-correction might become a target itself when one of its basic assumptions, such as uncorrelated errors, can be attacked.

\section{Conclusion}
\label{sec:conclusion}
Quantum machine learning is poised to deliver transformative gains in speed and accuracy for certain computational tasks, but these advantages rest upon equally transformative security foundations. In this work, we have explored the diverse threat landscape for QML by extending classical kill chain models to the quantum domain, highlighting how adversaries can exploit unique aspects of quantum hardware and software, from crosstalk side-channels and malicious transpiler modifications to data poisoning and backdoor insertion. By categorizing attacks and defenses across multiple stages of a QML adversarial lifecycle, we illustrate how security vulnerabilities often interlink, reinforcing the need for a holistic, defense-in-depth approach. Our analysis revealed both promising defenses (e.g., circuit equivalence checking, differential privacy for quantum data, noise-based robustness) and significant gaps, such as a general lack of threat modeling and notable research on multi-stage attacks.

Going forward, practical, large-scale applications of QML will require the community to adopt integrated countermeasures - spanning hardware-level isolation, verified transpilation, and robust quantum architectures - together with broader consensus on evaluation benchmarks and threat modeling frameworks. By systematically addressing these challenges, researchers, practitioners, and stakeholders can ensure that quantum machine learning systems can outperform classical models where it matters, but do so without compromising security.

\section*{Acknowledgment}
This research is part of the \emph{QML ESA} (Extended Security Analysis of Quantum Machine Learning) project, which is funded by the German Federal Office for Information Security (BSI).

\printbibliography

\end{document}